%% file: main.tex
\documentclass[conference]{IEEEtran}
\IEEEoverridecommandlockouts
\usepackage[utf8]{inputenc}
\usepackage{graphicx}
\usepackage{float} 
\usepackage{tikz}
\usepackage{xcolor}
\usepackage{caption}
\usepackage{subcaption}
\usepackage{hyperref}
\usepackage{placeins}
\usepackage{comment}
\usepackage{booktabs} 
\usepackage{multirow}


\usepackage{amsfonts}
\usepackage{pifont}
\newcommand{\cmark}{\textcolor{green!80!black}{\ding{51}}}
\newcommand{\xmark}{\textcolor{red}{\ding{55}}}

\definecolor{lime}{HTML}{A6CE39}
\DeclareRobustCommand{\orcidicon}{
	\begin{tikzpicture}
		\draw[lime, fill=lime] (0,0)
		circle[radius=0.16]
		node[white]{{\fontfamily{qag}\selectfont \tiny \textbf {\.{I}D}}};
	\end{tikzpicture}
	\hspace{-2mm}
}
\foreach \x in {A, ..., Z}{%
	\expandafter\xdef\csname orcid\x\endcsname{\noexpand\href{https://orcid.org/\csname orcidauthor\x\endcsname}{\noexpand\orcidicon}}
}

\definecolor{dodgerblue}{RGB}{64, 67, 194}
\newcommand{\Reviewed}{\color{black}}

\title{Federated Analytics for 6G Networks: Applications, Challenges, and Opportunities}
	\author{%
		\IEEEauthorblockN{
			\parbox{\linewidth}{\centering
				  Juan Marcelo Parra-Ullauri \hspace{-1.5mm}\orcidB{},\IEEEmembership{ Member,~IEEE}, Xunzheng Zhang\hspace{-1.5mm}\orcidA{},\IEEEmembership{Student Member,~IEEE}, \\ Anderson Bravalheri\hspace{-1.5mm}\orcidC{}, Shadi Moazzeni \hspace{-1.5mm}\orcidG{},\IEEEmembership{ Member,~IEEE},
				Yulei Wu\hspace{-1.5mm}\orcidD{},\IEEEmembership{ Senior Member,~IEEE}, \\
				Reza Nejabati\hspace{-1.5mm}\orcidE{},~\IEEEmembership{Senior Member,~IEEE}, and Dimitra Simeonidou\hspace{-1.5mm}\orcidF{},~\IEEEmembership{Fellow,~IEEE}%
			}%
		}%
		\thanks{The authors are with the High-Performance Networks Group, Smart Internet Lab, School of Computer Science, Electrical and Electronic Engineering, and Engineering Maths (SCEEM), Faculty of Engineering, University of Bristol, BS8 1QU, U.K. (e-mail: jm.parraullauri@bristol.ac.uk).}
	}

\date{June 2023}

\begin{document}

\maketitle

\begin{abstract}
Extensive research is underway to meet the hyper-connectivity demands of 6G networks, driven by applications like XR/VR and holographic communications, which generate substantial data requiring network-based processing, transmission, and analysis. However, adhering to diverse data privacy and security policies in the anticipated multi-domain, multi-tenancy scenarios of 6G presents a significant challenge. \emph{Federated Analytics (FA)} emerges as a promising distributed computing paradigm, enabling collaborative data value generation while preserving privacy and reducing communication overhead. FA applies big data principles to manage and secure distributed heterogeneous networks, improving performance, reliability, visibility, and security without compromising data confidentiality. This paper provides a comprehensive overview of potential FA applications, domains, and types in 6G networks, elucidating analysis methods, techniques, and queries. It explores complementary approaches to enhance privacy and security in 6G networks alongside FA and discusses the challenges and prerequisites for successful FA implementation. Additionally, distinctions between FA and Federated Learning are drawn, highlighting their synergistic potential through a network orchestration scenario.
\end{abstract}


\section{Introduction}
\label{intro}

\input{01_intro}

\section{Preliminaries and Related Work}
\label{preliminaries}
\input{02_preliminaries}

\section{Federated Analytics for 6G Networks}
\label{proposal}

\input{03_proposal}

\section{Example scenario: Cross-Domain Intelligent Orchestration}
\label{example}
\input{04_example}

\section{Research Challenges and Open Issues}
\label{challenges}
\input{05_challenges}

\section{Conclusion}
\label{conclusion}
\input{06_conclusion.tex}
\section*{Acknowledgements}
This work has been partially sponsored by the UKRI MyWorld Strength in Places Programme (SIPF00006/1), the UK GOV DSIT (FONRC) project REASON, and the EU H2020 5GASP (grant agreement No.~101016448).
\bibliographystyle{IEEEtran}
\bibliography{main} 
\newpage
\begin{IEEEbiographynophoto}{Juan Marcelo Parra-Ullauri} [M] obtained a bachelor's degree in engineering in Electronics and Telecommunications from the University of Cuenca, Ecuador in 2017. In 2023, he obtained a PhD in Computer Science from Aston University in the UK. Currently, he works as a Senior Research Associate at the Smart Internet Lab. His research interests include the Internet of Things, Distributed Machine Learning, Cloud Computing, Explainability in Autonomous Systems, and Data Engineering.
	\end{IEEEbiographynophoto}
 \vspace{-30pt}
\begin{IEEEbiographynophoto}
{Xunzheng Zhang}
		[S] received the B.Sc. and M.Sc. degree of Communication Engineering from Shandong University, China. He is currently pursuing the Ph.D. degree at the University of Bristol, UK. He has experience with Internet of Things and edge-cloud collaboration testbed deployment under Industry 4.0. He has the qualification of electronic technology application engineer, owns patents and won the Best Paper Award of IEEE International Symposium on Personal, Indoor and Mobile Radio Communications (IEEE PIMRC) 2020, London. His current research interests include 6G massive Internet of Things, federated learning, edge computing and machine learning.
	\end{IEEEbiographynophoto}
\vspace{-30pt}
 \begin{IEEEbiographynophoto}
{Anderson Bravalheri}
	received his Bachelor and Master degrees in Electrical Engineering
	(Telecommunications and Telematics) from University of Campinas (UNICAMP),
	Brazil in 2011 and 2016 respectively. From 2011 to 2017, he was a
	telecommunication researcher at the Research and Development Center in Telecommunications CPqD (Brazil). Currently, he is a
	Research Fellow in the Smart Internet Lab, University of Bristol, working
	with virtualistion and edge computing for advanced networks and immersive media
	applications. His research interests include distributed systems, fuzzy control,
	and AI solutions for networks.
	\end{IEEEbiographynophoto}
\vspace{-30pt}
 \begin{IEEEbiographynophoto}
 {Shadi Moazzeni}
 is a Research Fellow with the University of Bristol, Bristol, UK, where she is a member of the Smart Internet Lab and the High-Performance Networks research group. She received her M.Sc. degree in Computer Architecture Engineering from Amirkabir University of Technology (Tehran Polytechnic), Tehran, Iran in 2010, and her Ph.D. in Computer Architecture Engineering at the University of Isfahan, Iran in 2018. From July 2016 to February 2017, she was a Ph.D. visiting researcher at the University of Bologna, Italy. Her current research focuses on next generation intelligent network orchestration, multi-access edge computing, 6G networks intelligent profiling and orchestration towards zero-touch network and service management.
\end{IEEEbiographynophoto}
\vspace{-30pt}
 \begin{IEEEbiographynophoto}
{Yulei Wu}
        [SM] is an Associate Professor working across the Faculty of Engineering and the Bristol Digital Futures Institute, University of Bristol, UK. He is also affiliated with the Smart Internet Lab and is a member of the High Performance Networks Research Group. He received his Ph.D. degree in Computing and Mathematics and B.Sc. (1st Class Hons.) degree in Computer Science from the University of Bradford, UK, in 2010 and 2006, respectively. His research interests focus on digital twins and ethics-responsible decision making and their applications on future networks, connected systems, edge computing, and digital infrastructure.
	\end{IEEEbiographynophoto}
 \vspace{-30pt}
 \begin{IEEEbiographynophoto}{Reza Nejabati}
		[SM] is currently a chair professor of intelligent networks and head of the High-Performance Network Group in the Department of Electrical and Electronic Engineering in the University of Bristol, UK.  He is also a visiting professor and Cisco chair in the Cisco centre for Intent Based Networking in the Curtin University, Australia. He has established successful and internationally recognised experimental research activities in “Autonomous and Quantum Networks”. Building on his research, He co-founded a successful start-up company (Zeetta Networks Ltd) with 25 employees and £6m funding. His research received the prestigious IEEE Charles Kao Award in 2016 and has done important contributions in 5G, smart city, quantum communication, and future Internet experimentation.
	\end{IEEEbiographynophoto}	
 \vspace{-30pt}
 \begin{IEEEbiographynophoto}{Dimitra Simeonidou}
		[F] is a Full Professor with the University of Bristol, the Co-Director of the Bristol Digital Futures Institute and the Director of Smart Internet Lab. Her research is focusing in the ﬁelds of high performance networks, programmable networks, wireless-optical convergence, 5G/6G and smart city infrastructures. She is increasingly working with social sciences on topics of digital transformation for society and businesses. She has been the Technical Architect and the CTO of the Smart City Project Bristol Is Open. She is currently leading the Bristol City/Region 5G urban pilots. She has authored and coauthored over 600 publications, numerous patents, and several major contributions to standards. She has been the co-founder of two spin-out companies, the latest being the University of Bristol VC funded spin-out Zeetta Networks, delivering SDN solutions for enterprise and emergency networks.
		Prof. Simeonidou is a Fellow of Royal Academy of Engineering and a Royal Society Wolfson Scholar.
	\end{IEEEbiographynophoto}
	\vfill
\end{document}

%% file: 01_intro.tex
With 5G technology deployment and global standardisation, industry and academia efforts are dedicated to researching future 6G networks to meet projected demands~\cite{RQ6}. Societal digitisation, hyper-connectivity, and global data-driven ecosystems drive standardisation and strategic considerations for 6G networks to address communication requirements~\cite{RQ1}. Future 6G networks are expected to be more distributed, heterogeneous, intelligent, and closer to end-users~\cite{RQ6}. Achieving the ambitious 6G vision, including global coverage, diverse applications, robust security, spectrum utilisation, sensory integration, and full digitisation, requires further exploration of unresolved issues and research directions~\cite{RQ1}.

A key challenge in 6G networks arises from data-related issues due to advances in sensing, communication, and edge computing, resulting in a surge of data generation, transmission, and analysis within edge-cloud environments~\cite{wang2021federated}. Data is invaluable to Communication Service Providers (CSPs), with the top 50 carriers holding data from over five billion consumers~\cite{mangla2022application}. Research indicates data analytics will play a pivotal role in advancing network infrastructure towards 6G, offering rapid feedback, aiding troubleshooting, and converting raw data into actionable knowledge for automation systems~\cite{samdanis2023aiml}. Distributed analytics provide predictive capabilities, guiding algorithmic decisions to enhance system efficiency and infrastructure management, and can be integrated into architectural elements of 6G Network Data Analytics Functions (NWDAF)~\cite{zhou2023securing}.

The presence of heterogeneous data owners across diverse domains, multiple tenancies, and various technologies in 6G networks raises concerns about data privacy and confidentiality~\cite{wang2022federated}. The conventional edge-cloud computing paradigm, involving central server analysis, is insufficient for evolving application requirements~\cite{wang2021federated}. \emph{Federated Analytics (FA)} emerges as a promising distributed computing paradigm, enabling collaborative value generation from data across multiple remote entities while preserving local data for privacy and reduced communication overhead~\cite{elkordy2023federated}. FA uses big data principles and tools to manage and secure distributed and diverse data networks effectively, improving network performance, reliability, visibility, and security while preserving  confidentiality~\cite{wang2021federated}.

In this paper, our main focus is to showcase the benefits and applications of FA to tackle privacy and confidentiality concerns that arise from the diverse ownership at administrative, operational, and user levels, as well as the presence of various cloud and edge devices in 6G networks. We offer a thorough overview of FA in 6G networks, including its potential applications, domains, and types, as well as an exploration of data analysis methods, techniques, and queries. We also explore complementary approaches that enhance privacy and security when combined with FA. Furthermore, we discuss the open challenges for successful FA implementation in 6G networks. Lastly, we differentiate FA from Federated Learning (FL), exemplifying their synergistic potential through a network orchestration scenario.

The rest of this article is structured as follows. Section~\ref{preliminaries} explains 6G Networks, FA, and their distinctions from FL. We introduce a taxonomy for 6G FA and an implementation framework. We demonstrate the framework's working principles through an example and outline research challenges and open issues. In conclusion, we sum up this article.

%% file: 02_preliminaries.tex
\begin{figure*}
    \centering
    \includegraphics[width=4.8in]{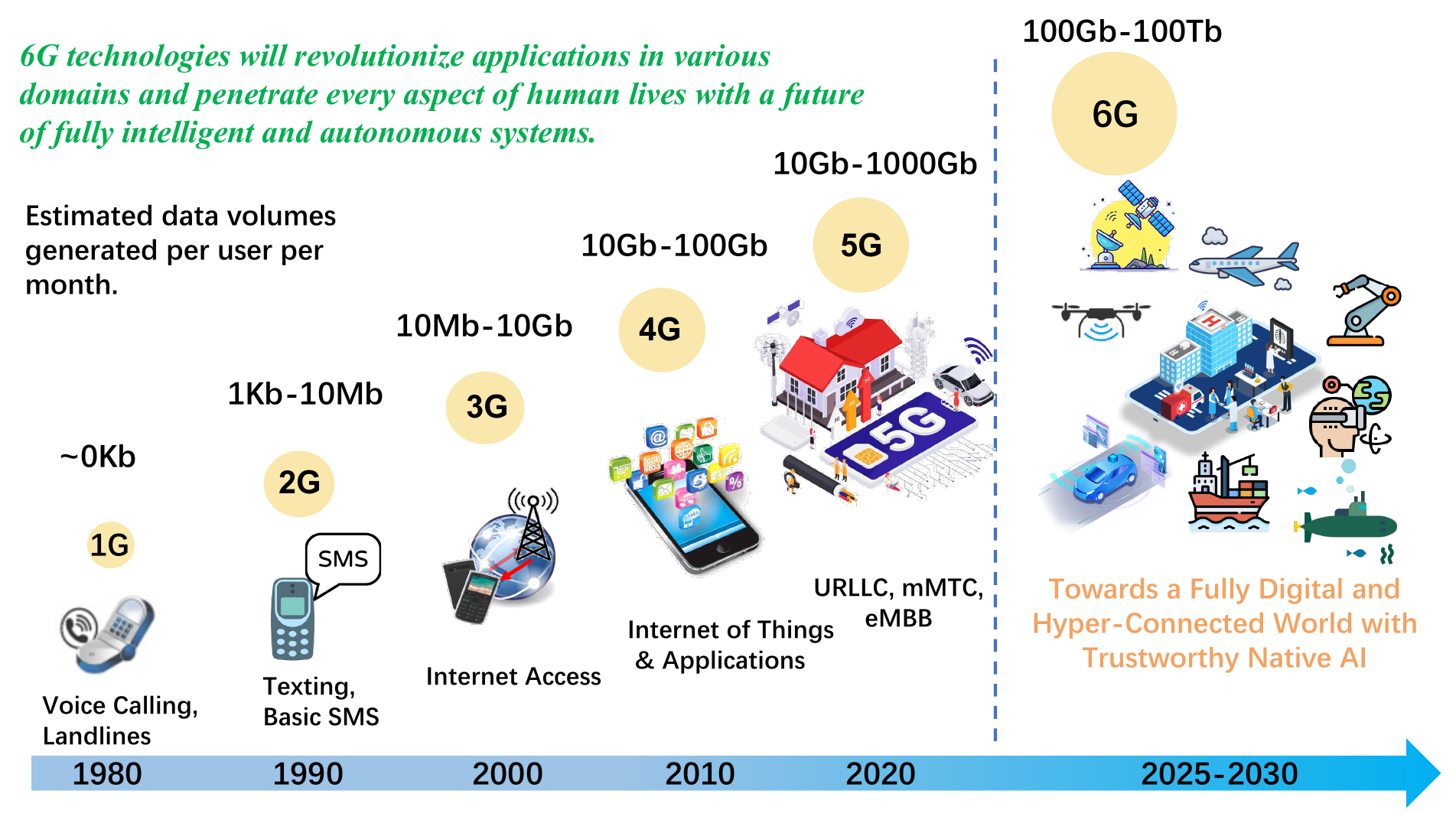}
    \caption{The estimated data volumes generated and evolution of networks towards future 6G.}
    \label{fig:6G}
\end{figure*}

\subsection{6G Networks}
The implementation of future services in 6G networks relies on seamless, instant, and virtually unlimited wireless connectivity~\cite{RQ1}. Set to materialize around 2030, 6G addresses challenges beyond the capabilities of backward compatibility with 5G NR evolution, attracting substantial attention from academia and industry with extensive research on 6G requirements and vision~\cite{RQ6}.

Key 6G requirements span technical aspects and application scenarios, encompassing data rate and delay with peak rates of up to 1 Tbps using cutting-edge technologies like THz and optical wireless communications~\cite{RQ1}. Additionally, factors such as user-experienced data rate, latency, security, and high mobility are critical in 6G systems~\cite{RQ6}. With an anticipated surge in data volumes (as showcased in Figure~\ref{fig:6G}), efficient data management, optimisation, privacy, and security are key in 6G networks~\cite{RQ1, RQ6}. The potential of 6G applications spans diverse domains, including IoT, smart grid 2.0, holographic telepresence, UAV-based mobility, extended reality, connected and autonomous vehicles, and intelligent healthcare~\cite{RQ6}. The integration of AI further accelerates 6G's evolution through scalable and trustworthy edge AI systems, decentralised ML models, AI of Things, and Haptic communication~\cite{RQ6}. As driving applications generate substantial data, privacy and confidentiality considerations become paramount~\cite{RQ6}. Proper evaluation of Key Performance Indicators (KPIs) will distinguish 6G networks from 5G counterparts in data rate, delay, capacity, coverage, service efficiency, and diversified service evaluation~\cite{RQ1} to meet the demands of an intelligent and hyper-connected society.

\subsection{Federated Analytics and Federated Learning}
FA is an emerging distributed knowledge aggregation paradigm designed to address data governance and privacy concerns related to data-sharing~\cite{elkordy2023federated}. FA conducts collaborative statistical analyses across multiple distributed nodes (referred to as \emph{FA clients}) without exchanging raw data among the parties~\cite{wang2021federated, wang2022federated}. Typically, FA involves a central querier (referred to as the \emph{FA server}) aiming to obtain knowledge or answer questions based on data distributed across various clients~\cite{elkordy2023federated}. FA clients receive the query, perform data-oriented tasks using their local data, and upload abstracted results to the server~\cite{wang2022federated}. In summary, as described in~\cite{elkordy2023federated}, FA's objective is for an FA server to respond to query $Q$ such as:
\begin{equation}
    Q(\mathcal{D}) = F_{\omega}(\mathcal{D}_{1}, \mathcal{D}_{2}, ... , \mathcal{D}_{N})
\end{equation}

where $\mathcal{D}: \{\mathcal{D}_i\}_{i=1}^n$ corresponds to the private raw data at the $N \in \mathbb{R}$ data owners (i.e., FA clients). $F_{\omega}$ refers to parameterised function describing the target query~\cite{wang2022federated}.     

Similar to FA, FL leverages decentralised data processing to address privacy concerns and enable collaboration among multiple entities. FL focuses on training ML models in a distributed privacy-preserving fashion~\cite{ullauri2023privacy}. In FL, the model is trained locally on the user's device or at the edge of the network, and only the model updates are exchanged with a central server or aggregator. The central server aggregates the updates from multiple devices and uses them to improve the global model~\cite{mangla2022application}. Considering the formulation of equation (1), FL can be viewed as a complex FA query on implemented at distributed data owners when the function $F_{\omega}$ describes an optimisation learning problem defined as:
\begin{equation}
    F_{\omega}(\mathcal{D}_{1}, \mathcal{D}_{2}, ... , \mathcal{D}_{N})= \arg\min_\omega \sum_{n=1}^N \frac{K_n}{K} \sum_{(x,y) \in \mathcal{D}_n} \ell(f(x;\omega), y)
\end{equation}

where $K$ represents the total number of data samples in all clients combined and $K_n$ represents the number of data samples in client $n \in N$. The main goal is to minimise for all clients the weighted sum of $\ell(f(x;\omega), y)$ that represents the loss function of model $f$ parameterised with $\omega$ which maps the discrepancy of predicted output and the true label $y$ for input $x$~\cite{wang2022federated,elkordy2023federated}.

Although FA and FL share federation characteristics, such as local model computation, central model aggregation, and interactive updates, they differ in their objectives and design details. While FL focuses on training ML models, FA is aimed at non-training data analytics tasks. As a result, the design details of FA vary due to the diverse nature of data analytics tasks~\cite{wang2022federated}. In 6G networks, FA and FL will play a crucial role for empowering intelligent nodes to learn and adapt locally, leading to reduced latency, improved network efficiency, and enhanced user experiences while maintaining privacy and will be explored in this document.  

\subsection{Related Work}
\newcommand{\tast}{\textasteriskcentered}
\newcommand{\tdag}{\textdagger}
\newcommand{\tddag}{\textdaggerdbl}

\begin{table*}[ht]
    \caption{Comparison$^\ast$ of studies about FA and Networks}
    \centering
\begin{tabular}{l c c c c c c c c}  
    \hline 
    \multicolumn{1}{c}{}                           & FA or FL & 6G     & Architecture & Methods & Applications & Types  & Domains & Privacy and Security\\
    \hline \hline 
    Wang et al.~\cite{wang2021federated}           & FA       & \xmark & \cmark       & \xmark  & \cmark       & \xmark & \xmark  & \cmark \\
    Zhao et al.~\cite{zhao2021communication}       & FA       & \xmark & \cmark       & \xmark  & \cmark       & \xmark & \xmark  & \cmark \\
    Wang et al.~\cite{wang2022federated}           & FA       & \xmark & \xmark       & \cmark  & \cmark       & \xmark & \cmark  & \xmark \\
    Samdanis et al.~\cite{samdanis2023aiml}        & FL       & \cmark & \xmark       & \cmark  & \xmark       & \cmark & subset  & \xmark \\
    Zhou et al.~\cite{zhou2023securing}            & FL       & \cmark & \cmark       & subset  & \cmark       & subset & subset  & \cmark \\
    Al-Quraan et al.~\cite{alquraan2023edgenative} & FL       & \cmark & \cmark       & subset  & \cmark       & subset & subset  & \cmark \\
    This work                                      & FA       & \cmark & \cmark       & \cmark  & \cmark       & \cmark & \cmark  & \cmark \\
    \hline
\end{tabular}
\vspace{1ex}
\label{tab:comparison}

\parbox{0.8\textwidth}{
    ${}^\ast$ FA methods, applications, types, domains and privacy and security concerns
    are described in detail in Section~\ref{sec:taxonomy}.
}

\end{table*}

In recent years, several works have emerged proposing the use of FA and related
techniques in networking and edge computing contexts.
Table~\ref{tab:comparison} summarises a few
important contributions and position our work with respect to the literature.

Wang et al.~\cite{wang2021federated} present a comprehensive overview of FA,
including a basic taxonomy and an abstract operating model, in addition to
proposing a high level architecture for FA that takes into consideration not
only privacy and security concerns but also basic resource and peer management
challenges.

Zhao et al.~\cite{zhao2021communication} propose a semihierarchical architecture for IoT data analytics, utilizing edge nodes as intermediary servers to aggregate models from local devices, improving convergence speed and reducing round-trip communications. They employ a random consensus algorithm for sharing model weights between servers. Wang et al.~\cite{wang2022federated} present a mechanism for performing FA in industrial IoT, considering data heterogeneity among clients. Skewness estimation for local data helps select a subset of clients for FA computations, resulting in improved accuracy and convergence.

Samdanis et al.~\cite{samdanis2023aiml} explore analytics in closed-loop automation systems' significance for next-gen network infrastructure. They propose a microservice-based AI and ML model architecture deployable in distributed or centralised modes. The study investigates inaccuracies due to data property changes (i.e., \emph{concept drift}) and reviews detection, prediction, and adaptation techniques. Zhou et al.~\cite{zhou2023securing} examine FL for implementing NWDAFs in 5G/6G. They propose a partial homomorphic encryption (HE) scheme for privacy and security preservation. A key management server generates and distributes encryption/decryption keys among FL participants. The scheme, validated in simulations, outperforms a prominent differential privacy algorithm without significant changes in prediction accuracy. Nevertheless, HE increases the computational time and resource requirements.

Al-Quraan et al.~\cite{alquraan2023edgenative} provide a comprehensive summary of FL techniques in 6G communications. Their systematic review covers enabling technologies, 6G drivers, and technical integration into network management systems. They compare the classical client-server architecture with a hierarchical client-edge-cloud FL approach and explore decentralisation alternatives like blockchain and peer-to-peer methods. The paper discusses client selection, incentivisation, data augmentation, and automatic labelling, addressing communication cost reduction, resource allocation, latency, and convergence time. The authors also outline privacy and security enhancement techniques for data sharing.

Although the
majority of works on FA (such~as~\cite{wang2021federated, zhao2021communication,
wang2022federated}), consider edge computing and some network-related
aspects, they don't specifically relate to either 5G or 6G
functionalities or take advantage of the infrastructure provided by these
platforms. In this paper, we review and expand on the taxonomy presented
in~\cite{wang2021federated} and explore the ramifications of FA from a
network-centric point of view, focusing on its implications for the development
of future 6G infrastructure.

On the other hand, works that explicitly target network infrastructure
management and orchestration or specific 5G/6G applications
(e.g.~\cite{samdanis2023aiml, zhou2023securing, alquraan2023edgenative}),
tend to consider forms of distributed analytics that
do not fully take advantage of the available FA techniques, or restrict
themselves to FL implementations. In this paper, we provide a comprehensive
review of different types and methods of FA that can be applied to 6G, covering
fundamental privacy and security aspects and outlining how these techniques can
be applied to support deployments in either single or multiple administrative
domains.










%% file: 03_proposal.tex
\subsection{An Architecure for Federated Analytics for 6G Networks}



\begin{figure}
    \centering
    \includegraphics[width=\columnwidth]{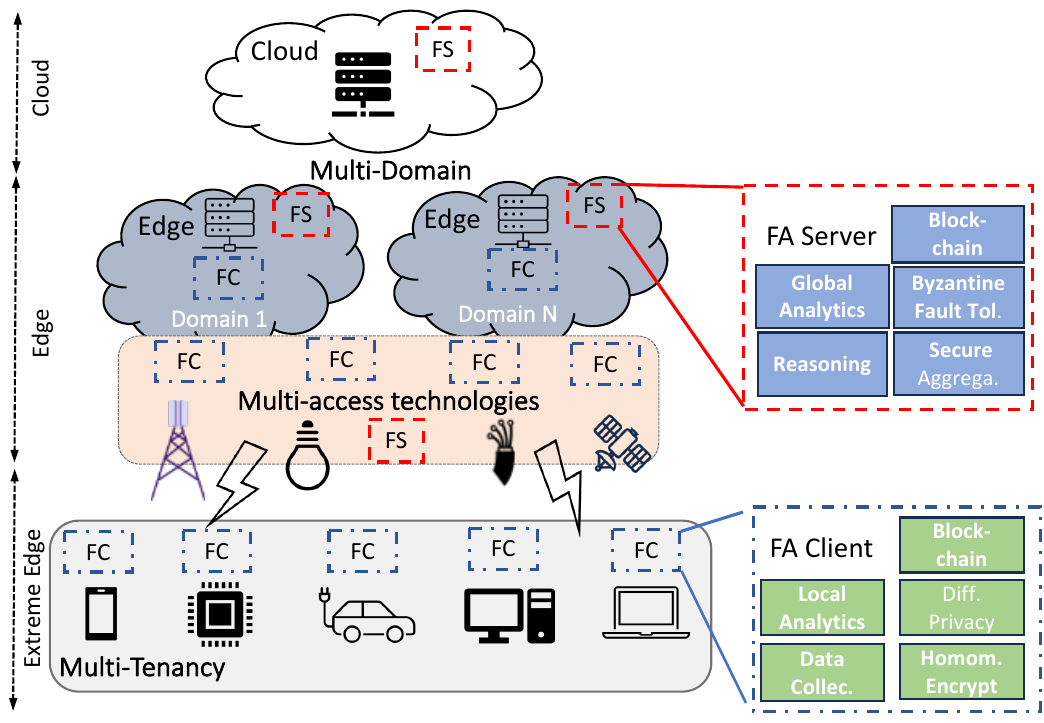}
    \caption{Federated Analytics for 6G Architecture}
    \label{fig:FA_arch}
\end{figure}

6G systems face privacy and confidentiality challenges due to distribution, heterogeneity, and diverse ownership at multiple levels. FA offers collaborative data analysis while maintaining privacy, with data decentralised across devices, reducing network data transfer. A vital component for FA in 6G is an architectural framework that integrates abstractions, supports functional segregation, and accelerates application development. Efficient peer administration, data structuring, and privacy governance are crucial to meet FA's unique demands. Figure~\ref{fig:FA_arch} depicts the proposed hierarchical framework. Analytics can be performed in each of the different distributed \emph{FA Clients (FC)} and then sent to central to be aggregated by the \emph{FA Server (FS)}.

In 6G networks, extreme edge scenarios envision multi-tenancy\footnote{We refer a tenant as the user owner of a task.} where users can request network infrastructure-as-a-service for task-offloading, introducing data that requires privacy and confidentiality handling. Each tenant can act as an FC, performing local data analytics to feed higher-level processes through an FS for resource monitoring and allocation, addressing privacy from an individual perspective.

At the edge level, 6G networks are expected to be highly autonomous, supporting different access technologies for a connected world. Multi-access and multi-administrative domain setups will involve heterogeneous services owned by different entities, requiring privacy-preserving FA at various levels for data confidentiality. Privacy concerns will be tackled from operational and administrative viewpoints.

In the cloud level, 6G networks will involve multi-administrative domain scenarios, where ensuring data privacy and confidentiality becomes complex as domains collaborate and share resources. Privacy concerns will be addressed operationally and administratively, developing robust mechanisms to protect data across diverse ownership structures and administrative boundaries. FA will play a vital role in enabling secure data sharing and processing, preserving privacy while facilitating higher-level insights and decisions.

The proposed framework aims to apply FA in 6G networks, addressing privacy concerns comprehensively from end-users to administrative and organisational domains.

\subsection{\Reviewed{Federated Analytics for 6G: A Taxonomy}}\label{sec:taxonomy}

\begin{figure*}
    \centering
    \includegraphics[width=0.55\textwidth, angle=270]{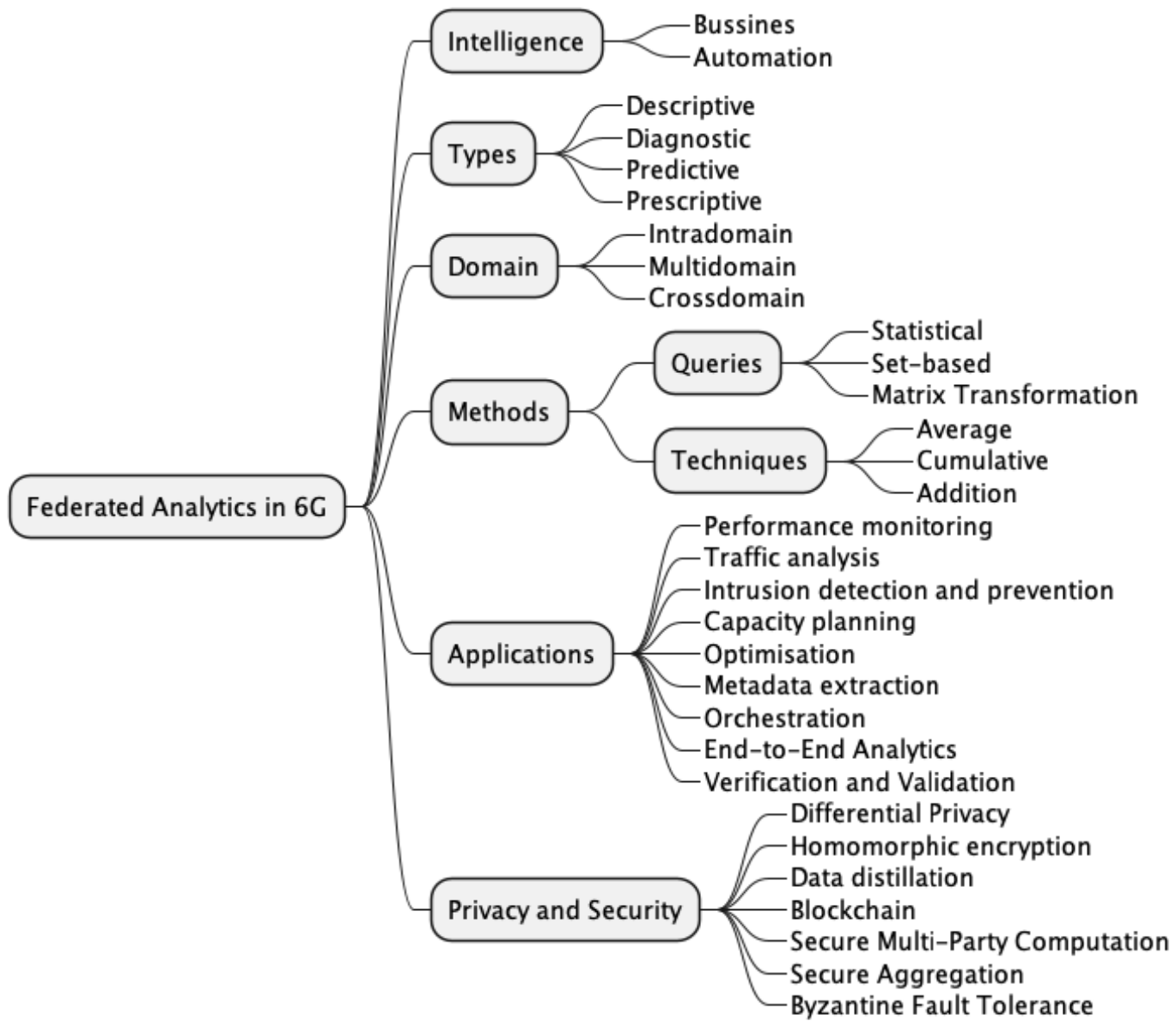}
    \caption{\Reviewed{Federated Analytics for 6G Taxonomy}}
    \label{fig:FA_tax}
\end{figure*}

This section presents a taxonomy (Figure~\ref{fig:FA_tax}) for classifying and organising concepts related to FA in 6G networks that can be exploited using the proposed architecture. 

\subsubsection{Intelligence}
The 5GPPP Architecture WG\footnote{\href{https://5g-ppp.eu/wp-content/uploads/2021/11/Architecture-WP-V4.0-final.pdf}{5GPPP View on 5G Architecture} accessed on 01/08/2023.} identifies various networking stakeholders, such as CSP and Virtual Service Providers (VSP), across different organisational domains with distinct business and operational responsibilities. Privacy concerns arise when these stakeholders use network data to enhance their operations, market understanding, and customer insights. FA in 6G networks allows stakeholders to access real-time data for privacy-aware Business Intelligence (BI) and Automation Intelligence to optimise decision-making, streamline operations, and automate tasks for resource allocation and innovative industry solutions.

\subsubsection{Types}
FA encompasses various data analysis techniques within a distributed network of data sources. FA types include: \emph{descriptive analytics} for understanding historical data and highlighting key summary statistics, \emph{diagnostic analytics} to identify patterns and reasons behind past events.

\emph{Predictive analytics} uses statistical and machine learning models to make future event predictions without centralizing sensitive data. \emph{Prescriptive analytics} goes further, offering data-driven recommendations for desired outcomes while respecting data privacy. FA allows collaborative analysis of distributed FC data, maintaining data security and decentralizing sensitive information.

\subsubsection{Domains}
In 6G networks, heterogeneous domains define administrative or functional boundaries for network segments, facilitating policy enforcement, security, and access control. \emph{Multi-domains} serve different organisational units, enabling tailored access control. \emph{Cross-domains} are managed by different organisations, supporting data exchange with autonomy and security. \emph{Intra-domains} are local segments under a single authority, enhancing communication and resource sharing.
FA benefits these domains in 6G networks. For multi-domains, FA enables collaborative data analysis and insights generation while preserving data privacy. In cross-domains, FA promotes secure data sharing and analysis, facilitating effective data collaboration. In intra-domains, FA enhances data analysis by pooling resources and knowledge, generating comprehensive insights and optimised resource allocation without centralising sensitive data.

\subsubsection{Methods}
In FA, various aggregation techniques combine data from different sources with data privacy and security. Key techniques include \emph{cumulative aggregation}, which accumulates values over a specific period or data points. \emph{Addition aggregation} involves adding values of the same data attribute across sources, calculating the total sum without revealing individual data points. Lastly, \emph{average aggregation} computes the mean or average value of a data attribute across sources, revealing overall averages without exposing data values.

An FA query (Eq. (1) in Sec.~\ref{preliminaries}) represents an FS question to distributed FCs, falling into three categories~\cite{elkordy2023federated}: a) \emph{Statistical testing queries} aim to discover statistical properties from private data. b) \emph{Set queries} focus on data associations like intersections and unions. c) \emph{Matrix transformation queries} include dimensionality reduction using methods like principal component analysis and projections.

\subsubsection{Applications}
In 6G networks, FA enables decentralised data processing and analysis, ensuring privacy and efficiency for decision-making among multiple entities. It supports applications such as \emph{metadata extraction}, where an FS queries an FC to gather local data insights, like the number of dataset classes and features. \emph{Performance monitoring} allows FCs to report ML model or service performance through local analytics. \emph{Verification and Validation} involve focused queries on local processes, giving FS a global system performance view without centralising raw data.

\emph{Intrusion and anomaly detection} benefit from FA's collaborative approach in network environments. Distributed intrusion detection systems collectively analyse data while preserving privacy, enhancing threat detection and secure local anomaly detection for improved network security. FA's knowledge extraction \emph{optimises} network resources and \emph{capacity planning} during \emph{orchestration} processes. Local analytics provide insights into process and infrastructure status, enabling \emph{end-to-end analytics} for data-driven decisions at the automation or business level.


\subsubsection{Privacy and Security} \label{sec:FA_6G_privacy_security}
are crucial in 6G, serving both end users and mission-critical vertical industries. FA supports network-wide data analytics without sharing local data, benefitting 6G in two key ways. Firstly, within various administrative domains (e.g., private and public networks, diverse edge device providers), FA enables analytics within each domain and consolidates insights for global results. Secondly, given 6G's multi-tenancy nature, data sharing between tenants is restricted. FA allows network management insights encompassing all tenants without data sharing, making it a promising paradigm for privacy-preserving data analytics in 6G.

However, there have been recent debates in FA regarding the potential privacy leakage of local data when sharing insights learned from local data in FA, despite the data not being shared outside its original region. Some existing approaches address this privacy issue, e.g., cryptography methods like \emph{HE} and \emph{secure multi-party computation}, privacy-enhancing technologies like \emph{differential privacy}, and \emph{secure aggregation}. Additionally, tampering with insights during FA's aggregation by adversaries can impact FA's performance, but this issue can be mitigated using technologies such as \emph{blockchain} and \emph{Byzantine Fault Tolerance}.

%% file: 04_example.tex
\begin{figure*}
    \centering
    \includegraphics[width=0.7\textwidth]{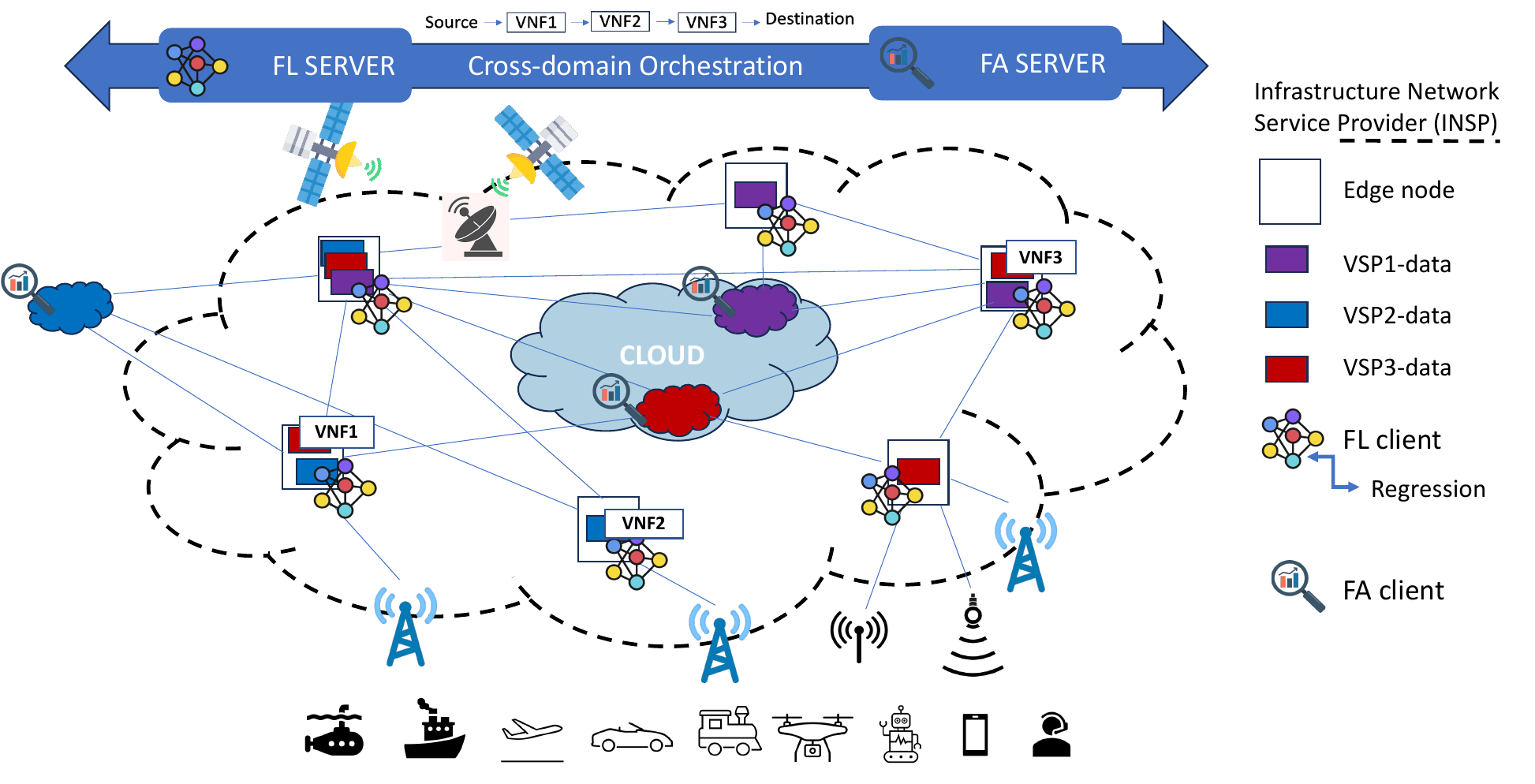}
    \caption{Federated Analytics and Learning for 6G Networks: Cross-Domain Profiling}
    \label{fig:FA_prof}
\end{figure*}

\begin{table*}
	\centering
	\caption{\\Federated Analytics: Statistical analysis without centralising datasets. VNF Firewall dataset from~\cite{moazzeni2020novel}.}
	\label{TabelFA}
	\begin{tabular} {c c c c c c c c c c c c c}
		\midrule
		&  &  &  &  & & \multicolumn{4}{c}{Distributed Datasets: Skewness per Feature} \\
		\cmidrule{4-12} 
	Client & N\_sam. & N\_feat. & CPUUTP & MEMUTP & RTT & MIR & CPU & MEM & In\_RX, & OuT\_TX & LINK & Selected? \\
		\midrule   
		1 & 895 & 9 & -0.7097 & 0.1806 & -0.5809 & 0.8565 & 0.3008 & -0.0123 & -0.0969 & -0.1929 & -0.0415 & yes\\
		2 & 400 & 9 & -0.7926 & 0.1822 & -0.4863 & 0.9215 & 0.3448 & 0.0325 & -0.0829 & -0.1818 & -0.0623 & yes\\
		3 & 100 & 9 & -1.0135 & 0.0411 & -0.3778 & 1.0952 & 0.4609 & 0.1288 & 0.039 & 0.1132 & 0.1107 & no\\
		4 & 120 & 9 & -0.6359 & 0.0419 & -0.6004 & 1.0823 & 0.2961 & -0.1357 & -0.1723 & -0.0749 & -0.216 & no\\
		5 & 400 & 9 & -0.6633 & 0.1731 & -0.5606 & 0.9464 & 0.3187 & 0.0603 & -0.0401 & -0.2112 & -0.1681 & yes\\
            6 & 330 & 9 & -0.7884 & 0.1024 & -0.5401 & 0.786 & 0.3227 & 0.0534 & -0.1275 & -0.1296 & -0.0588 & yes\\
            7 & 580 & 9 & -0.7906 & 0.1078 & -0.5066 & 0.8427 & 0.3451 & 0.0284 & -0.0409 & -0.1009 & -0.0317 & yes\\
            8 & 780 & 9 & -0.715 & 0.1878 & -0.6041 & 0.8647 & 0.2915 & -0.0274 & -0.1168 & -0.1914 & -0.0407 & yes\\
            9 & 500 & 9 & -0.666 & 0.2334 & -0.6553 & 0.8505 & 0.2909 & -0.0154 &-0.0884 & -0.0907 & -0.074 & yes\\
            10 & 290 & 9 & -0.8571 & 0.2951 & -0.4723 & 1.002 & 0.4308 & -0.1155 & -0.0466 & -0.2576 & -0.0057 & no\\
            \bottomrule
	\end{tabular}
 	\smallskip
        \smallskip
	\begin{tabular}{l}
	\end{tabular}
\end{table*} 
\subsection{Scenario}
Throughout this document, it's emphasised that 6G networks will have a vast array of heterogeneous devices and technologies from various operational and administrative domains, raising significant privacy concerns related to data ownership and confidentiality. In this context, FA and FL emerge as promising privacy-preserving distributed approaches, enabling valuable insights from local client data without centralising raw data, ensuring data security, and promoting collaborative intelligence.

An example of FA and FL application in 6G networks is cross-domain intelligent orchestration, involving the efficient allocation of computing resources across different domains. In this scenario (Figure~\ref{fig:FA_prof}), various VSPs utilise shared infrastructure with different access technologies from the INSP to provide services. Different SPs can share virtual and physical resources in 6G to meet service requirements, which can involve deploying various Virtual Network Functions (VNFs). The challenge is selecting the optimal host domain, edge nodes, and resources for deploying VNFs while adhering to VSP data ownership policies.

\begin{figure}
    \centering
    \includegraphics[width=0.9\columnwidth]{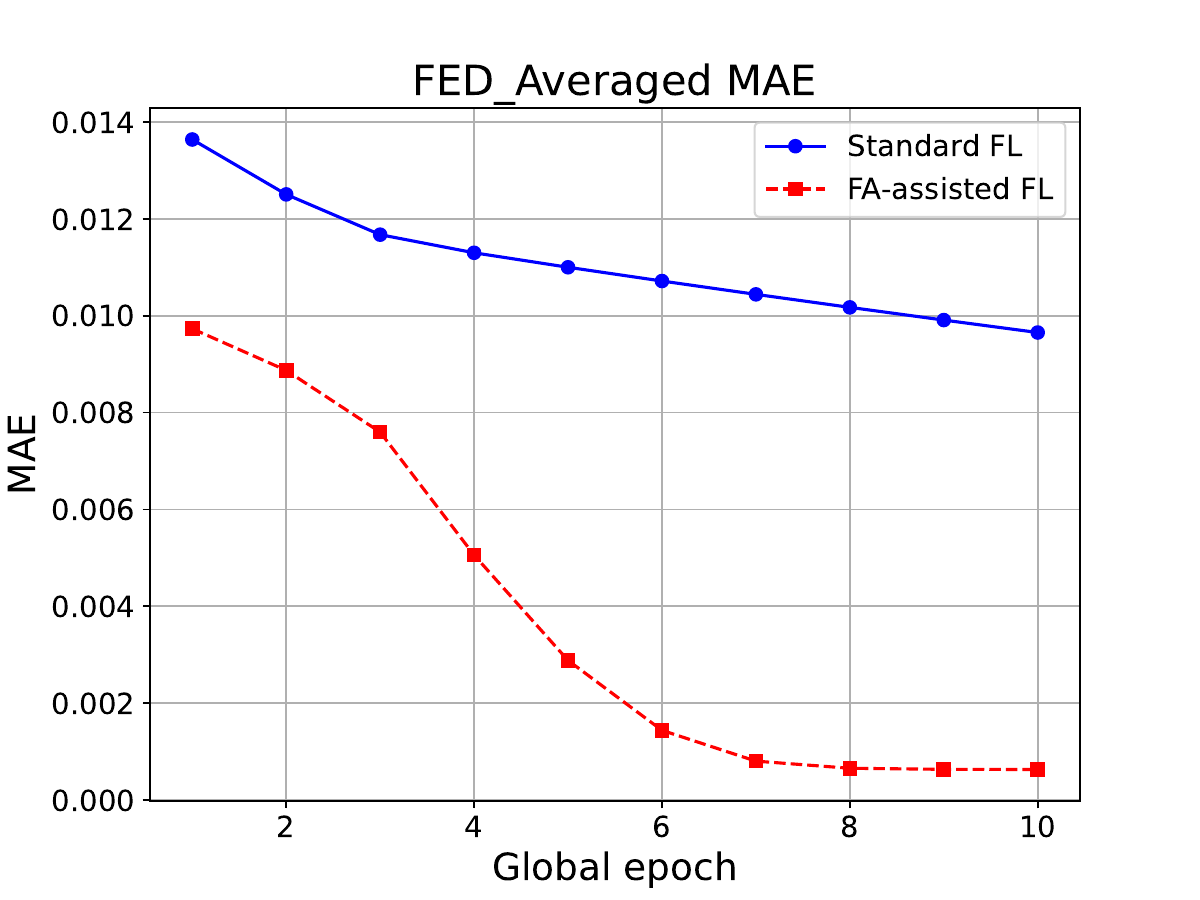}
    \caption{\Reviewed{MAE for predicting LINK capacity: Standard FL vs Proposed FA-assisted FL for VNF Firewall dataset from~\cite{moazzeni2020novel}.}}
    \label{fig:MAE}
\end{figure}

{\Reviewed{FA can enhance network management and resource optimisation while preserving data privacy. From the proposed architecture (Section~\ref{proposal}), an FS can be deployed at the cross-domain orchestrator level, which can be in the cloud. FCs are deployed in different virtual edge nodes across various VSPs on the INSP's physical edge nodes. Various FA queries $Q_i$ run over different edge nodes from different domains to gather knowledge about local resources and data availability for VNF profiling used in intelligent orchestration~\cite{moazzeni2020novel}. Statistical testing queries analyse FC dataset quality to determine eligible clients for joining an FL process to create VNF profiles, while set queries assess FC similarities and suitability for the FL system. Matrix transformation queries reduce dimensionality for selected clients before initiating the FL process. VNF profiles empower intelligent orchestration. Profiling predicts the maximum network load a VNF can handle and estimates required resources like CPU, memory, and network to meet performance targets and workloads~\cite{moazzeni2020novel}. This knowledge aids effective resource allocation and capacity planning while addressing privacy concerns in cross-domain orchestration.

\subsection{Experiment Definition and Evaluation}
We demonstrate the proposed approach's feasibility with a case study prototype implementation for VNF profiling in a Federated pipeline on a real network testbed based on the infrastructure setup from our previous work~\cite{ullauri2023privacy}.
\subsubsection{Dataset, model, and task}
We use the Virtual Firewall dataset from \cite{moazzeni2020novel}. This dataset is comprised of 9 features, 8 predictors and the target. The predictor variables are CPU utilization (CPUUTP), Memory utilization (MEMUTP), Network latency (RTT), VNF maximum input rate (MIR) and Packet loss (In\_RX, Out\_Tx), VNF resource configurations like CPU cores (CPU), Memory (MEM), and the target is the Link Capacity (LINK). We partition the dataset randomly to generate sub-datasets for each client. For this experiment, we use a simple
fully connected network with 3 hidden layers with 24, 12, and 6 neurons respectively. For the federation processes (i.e., FA and FL), we use the framework Flower built on top of the Kubernetes setup of~\cite{ullauri2023privacy}. The objective of the experiment is to use FA to examine the datasets of 10 clients scattered in the network. Based on this analysis, we select the clients that will be part of the FL process for cross-domain profiling. 
\subsubsection{Experimentation and results}

We use FA to conduct statistical data analysis on different edge nodes. We perform statistical queries to analyse data distribution and characteristics of clients without centralising the data. We specifically want to find the data points (Number of samples), the attributes (Number of features), and the skewness of the features in the individual FA clients’ datasets. Skewness is a measurement of the distortion of symmetrical distribution or asymmetry in a data set. Skewness can be quantified as a representation of the extent to which a given distribution varies from a normal distribution. The FA server accumulates the results into a matrix that is used for the analysis. This descriptive analysis allows us to choose the clients best suited for the FL process. 

Table~\ref{TabelFA} showcase the results of the FA process. We can observe that the different number of samples (N\_sam.) ranges between 100 and 895. The number of features (N\_feat.) is 9 for all clients. Regarding skewness, we can observe that none of the data points per feature follows a normal distribution. The distributions are either skewed to the right(+) or left(-). Based on this analysis, different approaches can be used for normalising the data distribution of the dataset. For example, data augmentation and balancing. However, in our experiment, we want to point out the importance of the data distribution for the FL process. We select clients with datasets with a greater number of samples ($<300$) and less skewed data distributions ($-1<x<1$). Based on these criteria, the clients selected for the FL are 1, 2, 4, 5, 6, 7, 8, and 9. 

We then assess the performance of standard FL compared to FA-enhanced FL. Figure~\ref{fig:MAE} showcase the results. It is demonstrated that by selecting clients based on their data distribution and characteristics, we enhance the performance of cross-domain profiling reducing the global Mean Absolute Error (MAE) loss for predicting the LINK capacity. Mapping back to our taxonomy of Section~\ref{proposal}, in this experiment, we showcase how to perform \emph{statistical testing queries} and \emph{accumulate} the results for \emph{descriptive analytics} to analyse each client dataset. Moreover, FL \emph{cross-domain} profiling can be categorised as \emph{predictive analytics} used for intelligent \emph{orchestration} aiming at \emph{automation intelligence}.  
} }

%% file: 05_challenges.tex
\subsection{\Reviewed{FA for 6G Challenges }}
\subsubsection{FA potential future applications in 6G networks}

{\Reviewed{When compared to previous telecom generations, 6G presents unique challenges due to new physical channel designs (higher frequency bands and transmission rates) and radio network architectures (e.g., ``cell-free" environments, smart surfaces, and mobile/temporary access points). These complexities demand advanced monitoring and analytics systems for control and management algorithms. 6G also shifts focus towards Key Value Indicators (KVIs) and Quality of Experience (QoE) metrics, requiring novel mathematical and computational models for estimation and quantification. FA is considered a crucial enabler for addressing these 6G challenges.

For example, regarding Ultra-Large-Scale MIMO~\cite{RQ1}, FA's resource management capabilities can assist in optimising resource allocation and scheduling, enhancing spectral efficiency by leveraging data from distributed antennas and UEs. Furthermore, combined with FL, FA can potentially serve for interference mitigation by sharing relevant information among base stations and coordinating beamforming strategies. In Integrated Sensing and Communication (ISAC)~\cite{RQ1}, FA can improve spectrum utilisation through data analysis from sensors and communication devices, assisting dynamic spectrum management. In the Open RAN landscape, FA  can enhance network optimisation and offer insights into RAN component performance, enabling self-configuration and optimisation. Together with FL, FA can support cooperative resource sharing among O-RAN operators. In the context of Reconfigurable Intelligent Surfaces (RIS)~\cite{RQ1}, FA can contribute to the analysis of environmental data to optimise surface deployment. When combined with FL, it enables collective learning and adaptation among these surfaces. In all the previously mentioned 6G scenarios FA combined with FL can enhance automation and intelligence with a main focus on privacy preservation. 

Nevertheless, 6G also imposes technical difficulties to the deployment of FA techniques. In particular, 6G high speeds and low latency in data transmission depend on the development of novel FA algorithms that are more efficient, more precise and more reliable. Moreover, the high level of distribution and diversity of 6G’ computing capabilities (considering not only edge and cloud computing but also connected devices) require new FA designs for increased robustness and hyper-scalability.

\subsubsection{FA in Goal/Task-oriented Communications}
Goal-oriented communications (also referenced as task-oriented commutations~\cite{taskim1}) involve Semantics of Information (SoI), capturing properties of information related to its ``goal" and ``purpose" in exchanges. These paradigms enable the design of more efficient systems that process and transmit only the most significant information, conserving energy and channel resources without sacrificing effectiveness in achieving communication goals~\cite{taskim1}. FA can potentially be used for semantics extraction to identify the meaning or context from unstructured data to understand its underlying concepts, relationships, and relevant information while maintaining data locally. The extracted semantics or insights can then be aggregated at a higher level without exposing sensitive details about individual data points. This ensures that privacy is maintained throughout the process, as the raw data never leaves the individual entities' boundaries. Goal-oriented and semantic communications are still in the early stages and the convergence with FA requires further research.  
}}
\subsection{Federated Data Analytics Challenges}
\subsubsection{Federated analytics with non-iid data}
Previous work about FL in non-iid data has been widely covered. However, in the 6G era,  data fragmentation will become more pronounced, as we know from Big Data to massive amounts of small-sized data~\cite{samdanis2023aiml}. How to use FA to deal with massive small-sized data and extract knowledge from them is a promising research area. In FedACS \cite{wang2022federated}, the authors give three modules: insights derivation, skewness estimation and client selection.  How to make clients generate insensitive insights about their local data, and make the server aggregates the insights from non-iid data is still under research. How to standardise the way to get data insight into FA is also a direction to be studied.

\subsubsection{Federated analytics with synthetic data}
Under the secure condition that the raw data should be kept locally, the FA algorithm can try to make clients generate synthetic data or similar data like GANs. By analysing this kind of virtual data, FA may get insights and inferences from different clients who are isolated from each other. The analysed results then be deployed to the training participants to have better system performance. For example, in~\cite{MULTI} proposed a learning framework using synthetic data generated by a GAN where the image styles are transferred from one domain into another domain. 
6G scenarios should consider the generated data to enhance the model performance and also try to realise self-data generation with self-model improvement for AI-network mutual enhancement.

\subsubsection{Federated analytics with incentive mechanisms}
In the 6G era, increased participation of high-quality clients in federated training poses challenges in determining when and how to engage in FL. Combining FA with incentive mechanisms allows clients to execute more effective strategies, improving machine learning model performance. Challenges include evaluating client contributions and attracting and retaining more participants. FA can assess final learning performance, estimate data quality, share information, calculate compensation, and generate revenue. FA incentive mechanisms also apply to scenarios like data trading and task-oriented computation \cite{taskim1}.

\subsection{Decentralised FA}

The usage of a centralising server to aggregate FA results can create a single
point of failure and undermine the security, resilience and scalability of network systems. Decentralised alternatives to FL, like \textit{gossip learning} (GL) and
\textit{blockcain powered FA} (BFA) have been proposed to address these limitations.
Privacy-preserving techniques (e.g., differential privacy) can be used to
mitigate the privacy leakage issue as discussed in
Section~\ref{sec:FA_6G_privacy_security} in both GL and BFA.

GL employs peer selection and communication protocols to exchange parameters
directly between participants, without significant degradation of training
performance or accuracy~\cite{hegedus2021gossip}.
Different flow control mechanisms can be implemented to affect how the
messages travel through the network (e.g., preventing congestion).
Since the availability of gossip protocols is decoupled from GL algorithms,
decentralised analytics architectures inspired by GL can also be developed.

In BFA, each client broadcasts its local analytics results
and aggregates the received results from other clients along with its own results.
In the case of FL, for each round of local training, clients additionally compete with
each other based on certain consensus mechanisms~\cite{alquraan2023edgenative}.
The centralised FA server is replaced by the peer-to-peer blockchain system and
the aggregation of local results is dealt with by the blockchain system, which
enhances reliability.
Blockchain also provides verification of local analytics results, strengthening
privacy and security and allowing malicious local results to be removed before aggregation. 

However, blockchain comes with remarkable overheads in terms of increasing energy
consumption, computation demands, computation delay and storage
which significantly degrades 6G system performance.
How to bring the computational overheads down to an acceptable level for 6G is a challenge.
This requires the selection of appropriate types of blockchain systems (e.g.,
consortium blockchain) with more efficient consensus algorithms (e.g., HotStuff
- developed by Facebook, with linear complexity of the authentication nodes).
It also requires joint optimisation by considering the computation requirements
of the blockchain system and the FA system, the elastic computation and
heterogeneous communication resources of the 6G system, and the (stringent)
requirements of 6G networks and applications.

%% file: 06_conclusion.tex
The presence of diverse data owners and edge devices adds complexity to data privacy and confidentiality concerns in 6G Intelligent Networks. In response, FA emerges as a promising distributed computing paradigm for collaborative value generation from data while maintaining privacy and reducing communication overheads. FA offers significant advantages in managing and securing distributed and heterogeneous data networks in 6G systems. This paper discusses FA principles and benefits, proposes an implementation framework for 6G networks, and identifies research challenges and open issues. Understanding the potential of FA and addressing these challenges will shape the future of 6G networks and meet the evolving intelligent communication requirements of our hyper-connected society.

%% file: main.bbl
\begin{thebibliography}{10}
\providecommand{\url}[1]{#1}
\csname url@samestyle\endcsname
\providecommand{\newblock}{\relax}
\providecommand{\bibinfo}[2]{#2}
\providecommand{\BIBentrySTDinterwordspacing}{\spaceskip=0pt\relax}
\providecommand{\BIBentryALTinterwordstretchfactor}{4}
\providecommand{\BIBentryALTinterwordspacing}{\spaceskip=\fontdimen2\font plus
\BIBentryALTinterwordstretchfactor\fontdimen3\font minus
  \fontdimen4\font\relax}
\providecommand{\BIBforeignlanguage}[2]{{%
\expandafter\ifx\csname l@#1\endcsname\relax
\typeout{** WARNING: IEEEtran.bst: No hyphenation pattern has been}%
\typeout{** loaded for the language `#1'. Using the pattern for}%
\typeout{** the default language instead.}%
\else
\language=\csname l@#1\endcsname
\fi
#2}}
\providecommand{\BIBdecl}{\relax}
\BIBdecl

\bibitem{RQ6}
S.~Dang, O.~Amin, B.~Shihada, and M.-S. Alouini, ``What should 6g be?''
  \emph{Nature Electronics}, vol.~3, no.~1, pp. 20--29, 2020.

\bibitem{RQ1}
C.-X. Wang, X.~You, X.~Gao, X.~Zhu, Z.~Li, C.~Zhang, H.~Wang, Y.~Huang,
  Y.~Chen, H.~Haas \emph{et~al.}, ``On the road to 6g: Visions, requirements,
  key technologies and testbeds,'' \emph{IEEE Communications Surveys \&
  Tutorials}, 2023.

\bibitem{wang2021federated}
D.~Wang, S.~Shi, Y.~Zhu, and Z.~Han, ``Federated analytics: Opportunities and
  challenges,'' \emph{IEEE Network}, vol.~36, no.~1, pp. 151--158, 2022.

\bibitem{mangla2022application}
U.~Mangla, ``Application of federated learning in telecommunications and edge
  computing,'' in \emph{Federated Learning: A Comprehensive Overview of Methods
  and Applications}.\hskip 1em plus 0.5em minus 0.4em\relax Springer, 2022, pp.
  523--534.

\bibitem{samdanis2023aiml}
K.~Samdanis, A.~N. Abbou, J.~Song, and T.~Taleb, ``{AI}/{ML} {Service}
  {Enablers} \& {Model} {Maintenance} for {Beyond} {5G} {Networks},''
  \emph{IEEE Network}, pp. 1--10, 2023, conference Name: IEEE Network.

\bibitem{zhou2023securing}
C.~Zhou and N.~Ansari, ``Securing {Federated} {Learning} {Enabled} {NWDAF}
  {Architecture} with {Partial} {Homomorphic} {Encryption},'' \emph{IEEE
  Networking Letters}, 2023, conference Name: IEEE Networking Letters.

\bibitem{wang2022federated}
Z.~Wang, Y.~Zhu, D.~Wang, and Z.~Han, ``Federated analytics informed
  distributed industrial iot learning with non-iid data,'' \emph{IEEE
  Transactions on Network Science and Engineering}, 2022.

\bibitem{elkordy2023federated}
A.~R. Elkordy, Y.~H. Ezzeldin, S.~Han, S.~Sharma, C.~He, S.~Mehrotra,
  S.~Avestimehr \emph{et~al.}, ``Federated analytics: A survey,'' \emph{APSIPA
  Transactions on Signal and Information Processing}, vol.~12, no.~1, 2023.

\bibitem{ullauri2023privacy}
J.~M.~P. Ullauri, L.~F. Gonzalez, A.~C. Bravalheri, R.~Hussain, X.~Vasilakos,
  I.~Vidal, F.~Valera, R.~Nejabati, and D.~Simeonidou, ``Privacy preservation
  in kubernetes-based federated learning: A networking approach,'' in
  \emph{IEEE International Conference on Computer
  Communications-Workshops(AIDTSP 2023)}, 2023.

\bibitem{zhao2021communication}
L.~Zhao, M.~Valero, S.~Pouriyeh, L.~Li, and Q.~Z. Sheng,
  ``Communication-efficient semihierarchical federated analytics in iot
  networks,'' \emph{IEEE Internet of Things Journal}, vol.~9, no.~14, pp.
  12\,614--12\,627, 2021.

\bibitem{alquraan2023edgenative}
M.~Al-Quraan, L.~Mohjazi, L.~Bariah, A.~Centeno, A.~Zoha, K.~Arshad,
  K.~Assaleh, S.~Muhaidat, M.~Debbah, and M.~A. Imran, ``Edge-{Native}
  {Intelligence} for {6G} {Communications} {Driven} by {Federated} {Learning}:
  {A} {Survey} of {Trends} and {Challenges},'' \emph{IEEE Transactions on
  Emerging Topics in Computational Intelligence}, vol.~7, no.~3, pp. 957--979,
  Jun. 2023, conference Name: IEEE Transactions on Emerging Topics in
  Computational Intelligence.

\bibitem{moazzeni2020novel}
S.~Moazzeni, P.~Jaisudthi, A.~Bravalheri, N.~Uniyal, X.~Vasilakos, R.~Nejabati,
  and D.~Simeonidou, ``A novel autonomous profiling method for the
  next-generation nfv orchestrators,'' \emph{IEEE Transactions on Network and
  Service Management}, vol.~18, no.~1, pp. 642--655, 2020.

\bibitem{taskim1}
J.~Shao, Y.~Mao, and J.~Zhang, ``Task-oriented communication for multidevice
  cooperative edge inference,'' \emph{IEEE Transactions on Wireless
  Communications}, vol.~22, no.~1, pp. 73--87, 2022.

\bibitem{MULTI}
M.~M. Rahman, C.~Fookes, M.~Baktashmotlagh, and S.~Sridharan, ``Multi-component
  image translation for deep domain generalization,'' in \emph{2019 IEEE Winter
  Conference on Applications of Computer Vision (WACV)}.\hskip 1em plus 0.5em
  minus 0.4em\relax IEEE, 2019, pp. 579--588.

\bibitem{hegedus2021gossip}
I.~Hegedűs, G.~Danner, and M.~Jelasity,
  ``\BIBforeignlanguage{en}{Decentralized learning works: {An} empirical
  comparison of gossip learning and federated learning},''
  \emph{\BIBforeignlanguage{en}{Journal of Parallel and Distributed
  Computing}}, vol. 148, pp. 109--124, Feb. 2021.

\end{thebibliography}
